\documentclass[aps,twocolumn,showpacs]{revtex4}
\usepackage{graphicx}
\usepackage{amsfonts}
\usepackage{amssymb}
\usepackage{amsbsy}
\usepackage{amsmath}
\usepackage{mathrsfs}
\usepackage{latexsym}
\usepackage{natbib}
\usepackage{bm}
\usepackage{subfigure}
\usepackage{color}

\def\kr{\kappa}

\def\ksg{\mathrm{\varkappa}}

\def\rs{r_s}
\def\rstar{r_{\star}}
\def\scriplus{\mathscr{I}^{+}}
\def\scriminus{\mathscr{I}^{-}}

\def\observerminus{\mathbb{O}^{-}}
\def\observerplus{\mathbb{O}^{+}}

\begin{document}

\title{New coordinates for a simpler canonical derivation of the Hawking effect}

\author{Golam Mortuza Hossain}
\email{ghossain@iiserkol.ac.in}

\author{Chiranjeeb Singha}
\email{cs12ip026@iiserkol.ac.in}

\affiliation{ Department of Physical Sciences, 
Indian Institute of Science Education and Research Kolkata,
Mohanpur - 741 246, WB, India }
 

\date{\today}

\begin{abstract}

In order to achieve a Hamiltonian-based canonical derivation of the Hawking 
effect, one usually faces multiple hurdles. Firstly, the spacetime foliation 
using Schwarzschild time does not lead to hyper-surfaces which are always 
spacelike. Secondly, the null coordinates which are frequently used in covariant 
approach, do not lead to a true matter Hamiltonian. Recently, an exact canonical 
derivation was presented using the so-called near-null coordinates. However, 
there too one faces the difficulty of having to deal with non-vanishing matter 
diffeomorphism generator as the spatial decomposition involves a non-zero shift 
vector. Here we introduce a new set of coordinates which allows one to perform 
an exact canonical derivation of Hawking effect without having to deal with 
matter diffeomorphism generator.

\end{abstract}

\maketitle

\section{Introduction}

An asymptotic future observer perceives thermal emission in a black hole 
spacetime when one considers quantum fields in such classical geometry. 
This phenomenon is known as the Hawking effect \cite{hawking1975}. Usually, a 
very large number of microstates are needed to understand thermal emission from 
a body. However, a classical black hole can be described by only few parameters 
in Einstein's general theory of relativity 
\cite{book:carroll,book:Schutz,Fulling:1989nb,book:wald}. So one expects that 
the study of Hawking effect in principle might allow one to understand the 
possible, yet unknown, quantum theory of gravity and significant efforts have 
been made to understand the Hawking effect in many different ways 
\cite{Singh:2014paa,Dray1985,Kawai:2013mda, book:parker,
Singleton:2011vh,Bhattacharya:2013tq, Singh:2013pxf,Lapedes:1977ip, 
Ho:2015fja,Jacobson:2012ei,PhysRevD.46.2486, Lambert:2013uaa, fredenhagen1990,
Jacobson:2003vx, Kiefer:2002fp, Traschen:1999zr, 
Chakraborty:2015nwa,Chakraborty:2017pmn,Carlip:2014pma, 
DEWITT1975295, Ford:1997hb, Hollands:2014eia,Padmanabhan:2009vy,
Fulling1987135,Hinton:1982, Parikh:1999mf,Visser:2001kq, Davies:1974th, 
Wald1975}.

In the canonical approaches to quantum gravity, one decomposes the spacetime 
into spatial hyper-surfaces labeled by a suitable time parameter. Consequently, 
in order to explore the techniques that are often employed in such canonical 
quantization framework, it is desirable to have a Hamiltonian-based canonical 
derivation of the Hawking effect. In such an approach, however one faces 
multiple hurdles. Firstly, the hyper-surfaces for fixed Schwarzschild time are 
not always spacelike \cite{Melnikov:2001ex,Melnikov:2002qd,Weinstein:2001kw} and 
consequently Hamiltonian dynamics is not well-posed in such coordinates. 
Secondly, in the standard derivation of the Hawking effect one needs to find 
the relation between the ingoing and outgoing massless field modes as seen by 
two asymptotic observers at the past and the future null infinity respectively
\cite{hawking1975}. These field modes follow null trajectory and are 
conveniently described using null coordinates. However, null coordinates do not 
lead to a true matter Hamiltonian that can describe the dynamics of these modes.

In order to overcome these difficulties, recently a set of near-null coordinates 
is introduced in \cite{Barman:2017fzh} which allows one to perform an exact 
canonical derivation of the Hawking effect. Firstly, these near-null coordinates 
lead to a non-trivial matter Hamiltonian which describes the dynamics of the 
field modes. Secondly, these coordinates being structurally closer to the null 
coordinates, allow one to follow similar methods which are employed for null 
coordinates. Nevertheless, the usage of these near-null coordinates leads to the 
off-diagonal terms in the spacetime metric. The corresponding spacetime 
decomposition involves both the \emph{lapse} function as well as a non-vanishing 
\emph{shift} vector. Consequently, the dynamics of field modes depends not just 
on matter Hamiltonian but also on the matter diffeomorphism generator.

This article is organized as follows. In the section 
\ref{Schwarzchild-spacetime}, we review the key aspects of a Schwarzschild black 
hole spacetime. Then we discuss the difficulties that one faces while using 
Schwarzschild time for space-time foliation. Subsequently, we introduce a new 
set of coordinates which allows an exact canonical derivation of the Hawking 
effect. The spacetime decomposition into spatial hyper-surfaces using these 
coordinates does not involve any shift vector. Therefore, the usage of these 
coordinates leads to a much simpler Hamiltonian-based derivation of the Hawking 
effect.

\section{Schwarzschild spacetime}\label{Schwarzchild-spacetime}

Let us consider a Schwarzschild spacetime which is formed at some finite 
past, possibly due to the collapse of a matter shell whose exact dynamics 
however is not important for understanding the Hawking effect. The invariant 
distance element in the Schwarzschild spacetime is given by
\begin{equation}\label{SchwarzschildMetric0}
ds^2 = - f(r) dt^2 + f(r)^{-1} dr^2 + r^2 d\theta^2 
+ r^2 \sin^2\theta d\phi^2  ~,
\end{equation}
where $f(r) = \left(1- r_s /r\right)$ and $r_s = 2 G M$ is the Schwarzschild 
radius. Throughout the paper, we use \emph{natural units} where $c=\hbar=1$. 
It is well-known that the Hawking effect is ultimately connected with the 
structure of the Schwarzschild metric in the $t-r$ plane. Therefore, for 
simplicity now onward we consider $1+1$ dimensional Schwarzschild spacetime 
with the metric $g_{\mu \nu}$ along with the invariant distance
\begin{equation}\label{SchwarzschildMetric1}
ds^2 =  g_{\mu \nu} dx^{\mu} dx^{\nu} = - f(r) dt^2 + f(r)^{-1} dr^2~.
\end{equation}
In order to represent the Hawking quanta, here we consider a minimally 
coupled massless scalar field $\Phi(x)$ whose dynamics is governed by the 
action 
\begin{equation}\label{ScalarActionFull}
S_{\Phi} = \int d^{2}x \left[ -\frac{1}{2} \sqrt{-g} 
g^{\mu \nu} \partial_{\mu}\Phi(x) \partial_{\nu}\Phi(x) \right] ~.
\end{equation}
We shall ignore the back-reaction of this scalar field on the spacetime 
metric as done also in the standard derivation of the Hawking effect  
\cite{hawking1975}.

\section{Canonical formulation}\label{Hamilton-formulation}

It turns out that the Schwarzchild time $t$ is not a good choice of time 
parameter for canonical formulation as the hyper-surfaces with a fixed 
Schwarzschild time $t$ are not always spacelike. We may easily see it from the 
expression $ds^2_{|dt=0} = f(r)^{-1} dr^2$ where hyper-surfaces for fixed 
Schwarzschild time are spacelike  when $r>r_{s}$ and timelike when $r<r_{s}$ 
\cite{Melnikov:2001ex,Melnikov:2002qd,Weinstein:2001kw}. In order to consider the spatial region only outside 
the horizon, usually one defines the so-called \emph{tortoise coordinate} 
$\rstar$ such that $d\rstar = f(r)^{-1} dr$. By choosing suitable constant of 
integration, $\rstar$ can be expressed as
\begin{equation}\label{TortoiseCoordinate}
\rstar = r + r_s \ln \left(r/r_{s} - 1\right)  ~.
\end{equation}
The domain of $\rstar$ being $(-\infty,\infty)$, it covers only a part of the 
full Schwarzschild spacetime and the corresponding metric becomes
\begin{equation}\label{SchwarzschildMetric}
ds^2 =  f(r) \left[ - dt^2 + d\rstar^2 \right] ~,
\end{equation}
which differs from  1+1 dimensional Minkowski metric by a conformal 
transformation.

\subsection{Null coordinates}

In the standard derivation \cite{hawking1975}, the Hawking effect is realized by 
computing the Bogoliubov transformation coefficients between the ingoing field 
modes that originate from the past null infinity ($\scriminus$) and the outgoing 
field modes that arrive at the future null infinity ($\scriplus$) respectively. 
For massless scalar field, these field modes follow null trajectories and are 
conveniently described using ingoing and outgoing null coordinates, defined as
\begin{equation}\label{NearNullCoordinatesMinus}
v = t + \rstar ~~;~~  u  = t - \rstar  ~.
\end{equation}
Subsequently, using these Bogoliubov coefficients, one computes the expectation 
value of number density operator corresponding to an observer near future null 
infinity in the  vacuum state corresponding to an observer near past null 
infinity. This expectation value turns out to be the same as the blackbody 
spectrum at the Hawking temperature. Therefore, these null coordinates play key 
roles even in the basic formulation of the Hawking effect in the covariant 
approach. However, these null coordinates do not lead to a true Hamiltonian for 
the matter field (\ref{ScalarActionFull}) that  can describe the field dynamics. 
Consequently, these null coordinates are not suitable for performing a 
Hamiltonian-based canonical derivation of the Hawking effect.

\subsection{Timelike and spacelike coordinates}

In order to perform an exact canonical derivation of the Hawking effect, a set
of near-null coordinates is introduced in Ref. \cite{Barman:2017fzh}. In 
particular, a timelike  coordinate $\tau_{-}$ and a spacelike coordinate 
$\xi_{-}$ used by an observer near the past null infinity $\scriminus$, referred 
to as the observer $\observerminus$, are given by 
\begin{equation}\label{NearNullCoordinatesMinus}
\tau_{-} = t - (1-\epsilon)\rstar~;~~ 
\xi_{-} = -t - (1+\epsilon)\rstar  ~,
\end{equation}
where the parameter $\epsilon$ is taken to be small and positive such that  
$\epsilon \ll 1$ which signifies the naming of these coordinates as `near-null'.
Similarly, one introduces another set of timelike coordinate $\tau_{+}$ and  
spacelike coordinate $\xi_{+}$ for an observer near the future null infinity  
$\scriplus$. These coordinates are given by
\begin{equation}\label{NearNullCoordinatesPlus}
\tau_{+} = t + (1-\epsilon)\rstar ~~;~~
\xi_{+} = -t + (1+\epsilon)\rstar ~,
\end{equation}
and the corresponding observer is referred to as the observer $\observerplus$.
We note that the domain of the coordinates $\tau_{\pm}$ and $\xi_{\pm}$ both 
are $(-\infty,\infty)$.

\subsubsection{Domain of the parameter $\epsilon$}\label{spatial}

The main motivation for choosing the parameter $\epsilon$ to be very small in 
Ref. \cite{Barman:2017fzh} was to keep these coordinates structurally `near' to 
the null coordinates so that one could employ similar methods as used for 
null coordinates. However, in general, any value of the parameter $\epsilon$ in 
the domain $0<\epsilon<2$ allows one to maintain the timelike and spacelike 
characteristics of the coordinates $\tau_{\pm}$ and $\xi_{\pm}$ respectively. 
Therefore, these coordinates can, in principle, be used for the study of the 
Hawking effect using canonical formulation in the entire allowed domain of 
$\epsilon$ which is not necessarily small. However, such coordinates would then 
loose their `near-null' characteristics. We note that for both the observers 
$\observerplus$ and  $\observerminus$, the $1+1$ dimensional Schwarzschild 
metric (\ref{SchwarzschildMetric}) can be expressed as
\begin{equation}\label{GeneralNewMetric}
ds^2 = \frac{f(r)}{4} \left[ -\alpha d\tau_{\pm}^2 + \beta d\tau_{\pm} 
d\xi_{\pm} + \gamma  d\xi_{\pm}^2  \right]~,
\end{equation}
where $\alpha = (2\epsilon+\epsilon^2)$, $\beta = 2(2-\epsilon^2)$ and 
$\gamma = (2\epsilon-\epsilon^2)$. For the small values of the parameter 
$\epsilon$ \emph{i.e.} $\epsilon \ll 1$, the parameter $\beta$ is 
non-vanishing. Therefore, if one foliates the spacetime into spatial 
hyper-surfaces by using the time variables $\tau_{\pm}$, the presence of the
\emph{off-diagonal} terms in the metric leads to non-vanishing shift vector. 
This in turns forces one to deal with the non-vanishing matter diffeomorphism 
generator \cite{Barman:2017fzh}.

\subsubsection{Parameter $\epsilon = \sqrt{2}$}

However, one may notice that the \emph{off-diagonal} terms in the metric 
(\ref{GeneralNewMetric}) vanishes identically for both observers if one chooses 
$\epsilon=\sqrt{2}$ which implies $\beta=0$. Then the corresponding metric 
becomes
\begin{equation}\label{metric}
ds^2 = \frac{f(r)}{4} \left[ -\alpha d\tau_{\pm} ^2 + \gamma d\xi_{\pm}^2 
\right] ~\equiv~ g^{\pm}_{\mu\nu}dx^{\mu}dx^{\nu} ~, 
\end{equation}
where $\alpha = 2(\sqrt{2} + 1)$ and $\gamma = 2(\sqrt{2} - 1)$. Clearly, if we 
use $\tau_{\pm}$  as time parameters with $\epsilon = \sqrt{2}$, then the 
foliation of the spacetime into spatial hyper-surfaces does not involve any 
shift vector.

\subsubsection{Relation between spatial coordinates $\xi_{-}$ and $\xi_{+}$}

In order to perform the canonical derivation of the Hawking effect, a key task 
is to find the relation between the spatial coordinates $\xi_{-}$ and $\xi_{+}$ 
which are used by the two asymptotic observers. Firstly, from the equations 
(\ref{NearNullCoordinatesMinus}, \ref{NearNullCoordinatesPlus}), we note that
\begin{equation}\label{xirstarRelation}
{d\xi_{-}} _{|\tau_{-}} = - 2 d {\rstar} _{|\tau_{-}} ~~,~~
{d\xi_{+}} _{|\tau_{+}} = 2 d {\rstar} _{|\tau_{+}} ~.
\end{equation}
However, we may emphasize here that  there was no black hole when the ingoing 
modes relevant for Hawking effect left the $\scriminus$ as seen by the observer 
$\observerminus$. So one should view the coordinates ($\tau_{-},\xi_{-}$) 
subject to the condition $r_s\rightarrow 0$ which implies $f(r)\rightarrow 1$ 
and  $\rstar\rightarrow r$. Now, using the metric (\ref{SchwarzschildMetric1}), 
one can calculate the non-vanishing Christoffel symbols given by
\begin{equation}\label{ChristoffelSymbols}
\Gamma^{t}_{tr} = \Gamma^{t}_{rt} = -\Gamma^{r}_{rr} = 
\frac{f'(r)}{2f(r)}  ~~;~~
\Gamma^{r}_{tt}  = \frac{1}{2} f(r) f'(r) ~.
\end{equation}
By introducing an affine parameter $\sigma$ along the null trajectories which 
are defined by $ds^2=0$, the geodesic equations can be expressed as
\begin{equation}\label{GeodesicEquation}
\frac{d}{d\sigma} \left(f(r) \frac{dt}{d \sigma}\right)  = 0  ~~,~~
\frac{d^2 r}{d \sigma^2} = 0  ~.
\end{equation}
The Eqns. (\ref{GeodesicEquation}) admit solutions for $r$ as
\begin{equation}\label{rsolution}
r = C\sigma + D ~,
\end{equation} 
where $C, D$ are constants of integration. Given affine transformations are of 
the form $\sigma\to\sigma' = C\sigma + D$, the coordinate $r$ can also be viewed 
as an affine parameter. We have mentioned that for the observer 
$\observerminus$, one should view the coordinates ($\tau_{-},\xi_{-}$) subject 
to the condition $r_s\rightarrow 0$. Now if we consider  a pivotal point 
$\xi^0_{-}$ on a constant $\tau_{-}$ hyper-surface with $r^0$ being the 
corresponding value of the radial coordinate then the Eqn. 
(\ref{xirstarRelation}) implies
\begin{equation}
\label{iv}
(\xi_{-}-\xi^0_{-})_{|\tau_-} = 2(r^0-r)_{|\tau_-}  ~,
\end{equation}
where $(\xi_{-}-\xi^0_{-})_{|\tau_-}$ to be viewed as the spatial 
separation between any two ingoing null rays which were at the locations
$\xi_{-}$ and $\xi^0_{-}$ respectively on the spatial hyper-surfaces 
labelled by the  time parameter $\tau_-$.

On the other hand, when the relevant outgoing modes for Hawking radiation arrive 
at $\scriplus$, as seen by the observer $\observerplus$, the black hole has 
already been formed. So if we consider  a pivotal point $\xi^0_{+}$ on a 
constant $\tau_{+}$ hyper-surface then using the Eqns. 
(\ref{TortoiseCoordinate}) and (\ref{xirstarRelation}) one can express the 
spatial separation between two given outgoing null rays along the  hyper-surface 
as
\begin{equation}\label{xiplusdiff}
(\xi_{+}-\xi_{+}^0){_{|\tau_{+}}} =  2(r-r^0)_{|\tau_{+}}
+2 r_s\ln\left(1+\frac{r-r^0}{r^0-r_{s}}\right)_{|\tau_{+}} ~.
\end{equation}
We have already shown that the coordinate $r$ along both ingoing and outgoing 
null trajectories can be considered as affine parameter. Therefore, using 
geometric optics approximation we can relate the spatial separations of the 
ingoing and the outgoing modes as
\begin{equation}\label{rr0relation}
(r-r^0)_{|\tau_{+}}  =  C' (r^0-r)_{|\tau_{-}}   ~,
\end{equation}
where $C'$ is some constant. Given this constant $C'$ does not affect the 
final result, then for simplicity we set this value to be unity. By choosing 
$\xi_{-}^0 = 2(r^0-r_{s})_{|\tau_{+}}$ and $\xi_{+}^0 = 
\xi_{-}^0 + 2r_{s}\ln\left(\xi_{-}^0/2r_s\right)$ in the Eqn. 
(\ref{xiplusdiff}), we can express it as
\begin{equation}\label{xiplusximinusrelation}
\xi_{+} = \xi_{-}  +  2 r_{s}\ln\left(\frac{\xi_{-}}{2r_s} \right) ~.
\end{equation}
In the domain where $|\xi_{-}|<<2r_{s}$, we may approximate the relation 
(\ref{xiplusximinusrelation}) between spatial coordinates $\xi_{-}$ and 
$\xi_{+}$ as used by two asymptotic observers $\observerminus$ and 
$\observerplus$ respectively, as
\begin{equation}\label{xirelationapprox}
 \xi_{-} \approx  2 r_{s} e^{\xi_{+}/2 r_{s}}~.
\end{equation}
The relation (\ref{xirelationapprox}) is the key relation which ultimately 
leads to the Hawking effect.

\begin{figure}
\includegraphics[width=8.5cm]{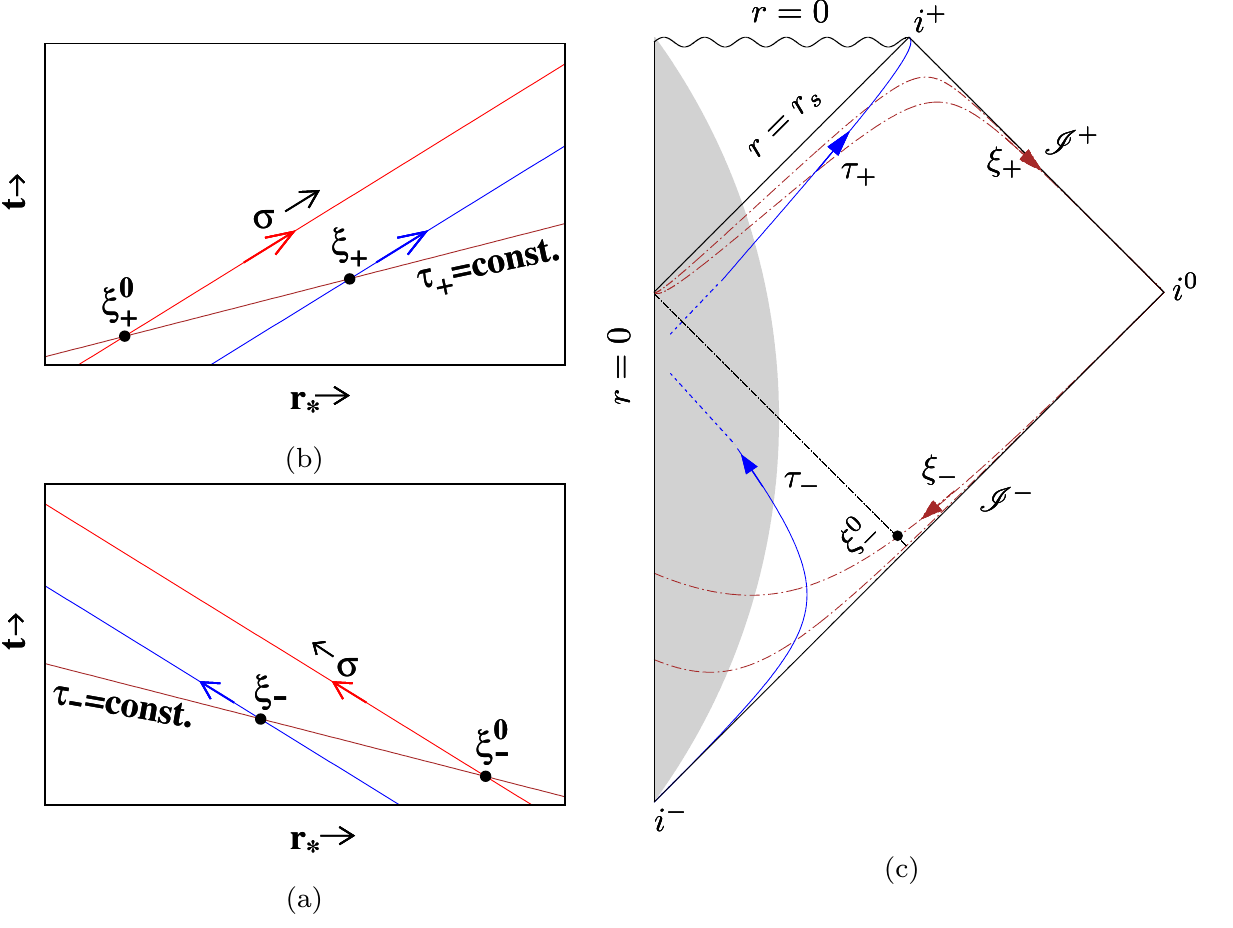}
\caption{(a)  Spatial separation between two ingoing null rays along a 
$\tau_{-}$ constant hyper-surface. (b) Spatial separation between two outgoing 
null rays along a $\tau_{+}$ constant hyper-surface. (c) The spacelike and 
timelike coordinates for $\epsilon=\sqrt{2}$ drawn on a Penrose diagram together 
with a collapsing shell of matter denoted by the shaded region.
}
\label{fig:CoordinatesPenrose} 
\end{figure}

\subsubsection{Scalar matter field}

We note that by using a conformally transformed spacetime metric 
$g^{0}_{\mu\nu}$ such that $g^{\pm}_{\mu\nu} = \tfrac{1}{4} \gamma 
f(r)~g^{0}_{\mu\nu}$, the scalar field action (\ref{ScalarActionFull}) for 
both the observers can be written in the form
\begin{equation}\label{ReducedScalarAction2DFlat}
S_{\varphi} =  \int d\tau_{\pm}  d\xi_{\pm} \left[-\frac{1}{2} \sqrt{-g^{0}} 
g^{0\mu\nu} \partial_{\mu}\varphi \partial_{\nu} \varphi \right]  ~,
\end{equation}
where the metric $g^{0}_{\mu\nu}$ is flat and consequently we can use the 
standard techniques of Fock quantization for the matter field. Using the time 
coordinates $\tau_{\pm}$, we can compute the scalar matter Hamiltonian as
\begin{equation}\label{ScalarHamiltonianFullMinus}
H^{\pm}_{\varphi} = \int d\xi_{\pm} ~ N \left[\frac{\Pi^2}{2\sqrt{q}}  + 
\frac{\sqrt{q}}{2}(\partial_{\xi_{\pm}}\varphi)^2 \right] ~,
\end{equation}
where the \emph{lapse function} $N = \sqrt{\alpha/\gamma} = (\sqrt{2}+1) $ and 
the determinant of the 
spatial metric $q = 1$. The Poisson bracket between the field $\varphi$ and 
its conjugate momentum $\Pi$ for both the observers can be expressed as
\begin{equation}\label{PoissonBracketMinus}
\{\varphi(\tau_{\pm},\xi_{\pm}), \Pi(\tau_{\pm},\xi_{\pm}')\} 
= \delta(\xi_{\pm} - \xi_{\pm}') ~.
\end{equation}
Using the equations of motion, the field momentum $\Pi$ can be expressed as 
\begin{equation}\label{FieldMomentumMinus}
\Pi(\tau_{\pm},\xi_{\pm}) = \frac{\sqrt{q}}{N} (\partial_{\tau_{\pm}}\varphi) ~.
\end{equation}

\subsubsection{Fourier modes}

The spatial volume $V_{\pm} = \int d\xi_{\pm}\sqrt{q}$ is formally 
divergent. Therefore, to avoid dealing with explicitly divergent quantity, we 
choose a fiducial box with finite volume as 
\begin{equation}\label{SpatialVoumeMinus}
V_{\pm} = \int_{\xi_{\pm}^L}^{\xi_{\pm}^R} d\xi_{\pm}\sqrt{q} = {\xi_{\pm}^R} - 
{\xi_{\pm}^L} \equiv L_{\pm} ~,
\end{equation}
where ${\xi_{\pm}^L}$ and ${\xi_{\pm}^R}$  are left and right coordinate edges 
associated with the box. We may now define the Fourier modes for the scalar 
field as \cite{Hossain:2010eb}
\begin{eqnarray}
\varphi(\tau_{\pm},\xi_{\pm}) &=&  \frac{1}{\sqrt{V_{\pm}}}\sum_{k} 
\tilde{\phi}^{\pm}_{k} 
(\tau_{\pm})~ e^{i k \xi_{\pm}}~,\nonumber\\
\Pi (\tau_{\pm},\xi_{\pm})&=&  \frac{1}{\sqrt{V_{\pm}}} \sum_{k} \sqrt{q}~ 
\tilde{\pi}^{\pm}_{k} (\tau_{\pm})~ 
e^{i k \xi_{\pm}}  ~,\label{FourierModesDefinitionMinus}
\end{eqnarray}
where complex-valued Fourier modes $\tilde{\phi}^{\pm}_{k}$ and 
$\tilde{\pi}^{\pm}_{k}$ are subject to the reality condition as we are 
considering the scalar field $\varphi$ to be a real-valued field. One may check 
that the Kronecker delta and the Dirac delta can now be expressed as
\begin{eqnarray}
\int d\xi_{\pm}\sqrt{q} ~ e^{i(k-k')\xi_{\pm}} = V_{\pm} \delta_{k,k'} 
~,\label{KroneckerDeltasMinus}\\
\sum_k e^{ik(\xi_{\pm}-\xi_{\pm}')} = V_{\pm} \delta(\xi_{\pm}-\xi_{\pm}')/\sqrt{q} 
~.\label{DiracDeltasMinus}
\end{eqnarray}
The Eqns. (\ref{KroneckerDeltasMinus}) and  (\ref{DiracDeltasMinus}) together 
allow the values of the wave-vector to be $k \in \{k_l ~| k_l = 2\pi 
l/L_{\pm}\}$ with $l$ being a non-zero integer. Using Fourier modes, 
the scalar field Hamiltonian (\ref{ScalarHamiltonianFullMinus}) for both the 
observers can be expressed  as $ H^{\pm}_{\varphi} = \sum_k N 
\mathcal{H}_k^{\pm}$ where the Hamiltonian density for the $k^{th}$ mode is
\begin{equation}\label{FourierHamiltonianDensity}
\mathcal{H}_k^{\pm} = \frac{1}{2} \tilde{\pi}^{\pm}_{k}  \tilde{\pi}^{\pm}_{-k}
+ \frac{1}{2} |k|^2 \tilde{\phi}^{\pm}_{k}  \tilde{\phi}^{\pm}_{-k} ~.
\end{equation}
The Poisson bracket between the Fourier modes and their conjugate momenta  
can be expressed as
\begin{equation}\label{FourierPoissonBracketMinus}
\{\tilde{\phi}^{\pm}_{k}, \tilde{\pi}^{\pm}_{-k'}\} = \delta_{k,k'} ~.
\end{equation}

\subsubsection{Relation between Fourier modes}

In order to establish the relation between the Fourier modes of two asymptotic 
observers, firstly we note that the matter field being scalar, it can be
expressed in general as 
$\varphi(\tau_{-}(\tau_{+},\xi_{+}),\xi_{-}(\tau_{+},\xi_{+})) = 
\varphi(\tau_{+},\xi_{+})$. Further, in the standard formulation of the Hawking 
effect, the observer near the $\scriminus$, deals with the \emph{ingoing} field 
modes for them $v = t + \rstar = (\tau_{-} - (\sqrt{2}-1)\xi_{-})/\sqrt{2}$  
is \emph{constant}. On the other hand, the observer near $\scriplus$ deals with 
the \emph{outgoing} field modes for them $u = t - \rstar = (\tau_{+} - 
(\sqrt{2}-1)\xi_{+})/\sqrt{2}$ is \emph{constant}. This aspect allows 
one to get a relation between the field momenta \cite{Barman:2017fzh} as
\begin{equation}
 \Pi(\tau_{+},\xi_{+}) = (\partial \xi_{-}/\partial \xi_{+}) 
\Pi(\tau_{-},\xi_{-})~.\nonumber
\end{equation}
The Fourier modes and the conjugate momenta on a given hyper-surface 
labeled by $\tau_{+}^0$, as seen by the observer $\observerplus$, can be 
expressed using the modes corresponding to the observer $\observerminus$,
on a given hyper-surface labeled by $\tau_{-}^0$, as
\begin{eqnarray}
\tilde{\phi}^{+}_{\kr}(\tau_{+}^0) &=& \sum_{k} \tilde{\phi}^{-}_{k}(\tau_{-}^0) 
F_{0}(k,-\kr) 
~,\label{FieldModesRelation}\\
\tilde{\pi}^{+}_{\kr}(\tau_{+}^0) &=&  \sum_{k} \tilde{\pi}^{-}_{k} 
(\tau_{-}^0)F_{1}(k,-\kr)  
~,\label{FieldMomentaModesRelation}
\end{eqnarray}
where the coefficient functions $F_{m}(k,\kr)$ are given by
\begin{equation}\label{FFunctionGeneral}
F_{m}(k,\kr) = \frac{1}{\sqrt{V_{-} V_{+}}} 
\int d\xi_{+} \left(\frac{\partial \xi_{-}}{\partial \xi_{+}} \right)^m
~e^{i k \xi_{-} + i \kr \xi_{+}} ~,
\end{equation}
with $m=0,1$. The coefficient functions $F_{m}(k,\kr)$ play the similar role 
as the Bogoliubov coefficients. Using the expression (\ref{FFunctionGeneral}), 
it can be shown that $F_{0}(k,\kr)$ and $F_{1}(k,\kr)$ are related as 
\cite{Barman:2018ina} 
\begin{equation}\label{F0F1Relation}
F_{1}(\pm|k|,\kr) = \mp \frac{\kr}{|k|}~F_{0}(\pm|k|,\kr) ~.
\end{equation}
The coefficient function $F_{0}(k,\kr)$ is formally divergent as the integrand 
is purely oscillatory. However, it can be evaluated by introducing a suitable 
regulator $\delta$ such that $\lim_{\delta\to 0} F_{0}^{\delta}(\pm|k|,\kr) = 
F_{0}(\pm|k|,\kr)$ and the regulated coefficient function can be evaluated as
\cite{Hossain:2014fma,Barman:2017fzh}
\begin{equation}\label{FFunctionEvaluated}
F_{0}^{\delta}(\pm|k|,\kr) = \frac{(2\rs)^{-\beta} |k|^{-\beta-1} }
{\sqrt{V_{-} V_{+}}}  e^{\pm i \pi(\beta+1)/2} ~\Gamma(\beta+1) ~,
\end{equation}
where $\Gamma(\beta+1)$ is the Gamma function and $\beta = (2i\kr\rs + \delta - 
1)$. From the Eqn. (\ref{FFunctionEvaluated}), one can deduce an important 
relation as follows
\begin{equation}\label{F0F0Relation}
F_{0}^{\delta}(-|k|,\kr) = 
e^{2\pi\rs\kr-i\delta\pi} ~F_{0}^{\delta}(|k|,\kr)  ~.
\end{equation}

\subsubsection{Number density of Hawking quanta}

Using the Eqns. (\ref{FieldModesRelation}), (\ref{FieldMomentaModesRelation}), 
(\ref{F0F1Relation}) and (\ref{F0F0Relation}) one can express the Hamiltonian 
density (\ref{FourierHamiltonianDensity}) corresponding to the \emph{positive} 
frequency modes \emph{i.e.} $\kr>0$ for the observer $\observerplus$ in terms 
of the Fourier modes of the observer $\observerminus$ as \cite{Barman:2017fzh}
\begin{equation}\label{ModesHamiltonianRelations0}
\frac{\mathcal{H}_{\kr}^{+}}{\kr} = \frac{h_{\kr}^1}{\kr}+ 
\frac{e^{2\pi\kr/\ksg} + 1}{e^{2\pi\kr/\ksg} - 1} \left[ 
\frac{1}{\zeta(1+2\delta)} \sum_{l=1}^{\infty} \frac{1}{l^{1+2\delta}} ~ 
\frac{\mathcal{H}_{k_l}^{-}}{k_l} \right]~,
\end{equation}
where $\ksg=1/(2\rs)$ is the \emph{surface gravity} at the Schwarzschild event 
horizon and $\zeta(1+2\delta) = \sum_{l=1}^{\infty} l^{-(1+2\delta)}$ is the 
\emph{Riemann zeta function}. The term $ h_{\kr}^1 =  \sum_{k\neq k'} [ 
\frac{1}{2} F_{1}(k,-\kr) F_{1}(-k',\kr) 
~ \tilde{\pi}^{-}_{k}  \tilde{\pi}^{-}_{-k'} + \frac{1}{2} |\kr|^2 F_{0}(k,-\kr) 
F_{0}(-k',\kr) ~ \tilde{\phi}^{-}_{k} \tilde{\phi}^{-}_{-k'}]$ being linear
in Fourier modes and their conjugate momenta, would drop out from the vacuum 
expectation value. 
It is well known that the Fourier modes corresponding to a massless free scalar 
field can be viewed as a system of decoupled harmonic oscillators which can also 
be seen from the Eqn. (\ref{FourierHamiltonianDensity}). Therefore, in Fock 
quantization $\langle\hat{\mathcal{H}}_{k}^{-}\rangle \equiv \langle 0_{-}| 
\hat{\mathcal{H}}_{k}^{-}|0_{-}\rangle = \frac{1}{2}|k|$ where the state 
$|0_{-}\rangle$ refers to the vacuum state of the observer $\observerminus$. 
Consequently, the expectation value of the number density operator 
$\hat{N}^{+}_{\kr} \equiv \hat{\mathcal{H}}_{\kr}^{+}/\kr - \frac{1}{2} $ 
corresponding to the observer $\observerplus$, in the vacuum state of the 
observer $\observerminus$ can be evaluated as
\begin{equation}\label{NumberOperatorVEV}
N_{\omega} \equiv \langle \hat{N}^{+}_{\omega=\kr}\rangle = 
\frac{1}{e^{2\pi\omega/\ksg} - 1} = \frac{1}{e^{(4\pi\rs)\omega} - 
1} ~.
\end{equation}
The Eqn. (\ref{NumberOperatorVEV}) corresponds to a thermal spectrum of bosons 
at the temperature $T_H = \ksg/(2\pi k_B) = 1/(4\pi\rs k_B)$. This phenomenon is 
referred to as the Hawking effect and associated temperature is known as the 
Hawking temperature.

\bigskip

\section{Discussions}\label{discussion}

In this article we have presented an exact analytical derivation of the Hawking 
effect in canonical formulation where one does not need to deal with the matter 
diffeomorphism generator. In order to achieve this simplification, we have 
introduced a new set of coordinates in which the resultant spacetime metric is 
diagonal. Consequently, the foliation of the spacetime into spatial 
hyper-surfaces, which is required for canonical derivation, does not introduce 
any shift vector. Therefore, these new coordinates lead to a much simpler 
canonical derivation of the Hawking effect compared to the one reported in Ref. 
\cite{Barman:2017fzh} where one uses the so-called near-null coordinates. 
Clearly, these coordinates would be quite useful for testing various new 
quantization techniques \cite{Ashtekar:2002sn,HALVORSON200445,Hossain:2010eb, 
Hossain:2014fma, Hossain:2016klt, Hossain:2015xqa,Barman:2017vqx}. We have 
mentioned earlier that the spacetime metric is diagonal in these new coordinates 
and up to a scaling the metric is similar to a conformally transformed Minkowski 
metric. However, it can be checked that these new coordinates cannot be obtained 
simply by applying a Lorentz boost from $(t,\rstar)$ coordinates. In this 
context we may mention that it would be quite interesting to use the canonical 
formulation as given here, to study the issue of ambiguity in the expression of 
Hawking temperature due to inequivalent choices of the inertial frames as shown 
by 't Hooft \cite{THOOFT198445,tHooft:1984kcu,Akhmedov:2006pg,Akhmedov:2008ru}.

\begin{acknowledgments}
We would like to thank Gopal Sardar, Subhajit Barman and Saumya Ghosh for many useful 
discussions. CS would like to thank IISER Kolkata for supporting this work 
through a doctoral fellowship. 
\end{acknowledgments}

\end{document}